\documentclass[twocolumn]{article}

\usepackage[T1]{fontenc}
 
\iffalse
\usepackage[scaled=0.92]{helvet}
\usepackage{mtpro2}
\ifdefined\nopdfsync\else\usepackage{pdfsync}\fi 
\else
\advance\textheight by 3cm
\advance\topmargin -2cm
\advance\textwidth by 2.4cm
\advance\oddsidemargin by -1.2cm
\advance\evensidemargin by -1.2cm
\fi

\usepackage[latin1]{inputenc}
\usepackage{graphicx}
\usepackage{amsfonts}
\usepackage{amsmath}
\usepackage{mathrsfs,theorem}
\usepackage{rotating}
\usepackage{algorithm, algorithmic}
\floatname{algorithm}{Algorithm}
\usepackage[usenames,dvipsnames]{color}
\usepackage{multirow}
\usepackage{dcolumn}

\usepackage{graphicx}
\usepackage{mathrsfs}
\usepackage{booktabs}
\usepackage{xspace}
\usepackage{algorithmic}
\usepackage{algorithm}
\usepackage{url}
\usepackage{algorithm, algorithmic}
\floatname{algorithm}{Algorithm}

\newfloat{Algorithm}{htpb}{alg}
\newcounter{prgline}

\newcommand{\Var}[1]{\mathrm{Var}[#1]}
\newcommand{\StdDev}[1]{\sqrt{\Var{#1}}}
\newcommand{\Nhat}{\hat N}

\renewcommand{\epsilon}{\varepsilon}

\def\..{\,\mathpunct{\ldotp\ldotp}} % Middle stuff for intervals. Usage: \..

\title{Four Degrees of Separation}

\iffalse
\numberofauthors{5}
\author{%
\alignauthor Lars Backstrom\\ \affaddr{Facebook}\\
	\email{lars@fb.com} 
\alignauthor Paolo Boldi\\ % \affaddr{Dipartimento di Scienze dell'Informazione}\\
	\affaddr{DSI, Universit\`a degli Studi di Milano, Italy}\\
	\email{boldi@dsi.unimi.it} 
\alignauthor Marco Rosa \\
	\affaddr{DSI, Universit\`a degli Studi di Milano, Italy}\\
	\email{marco.rosa@unimi.it} \and
\alignauthor Johan Ugander\\ \affaddr{Facebook}\\
	\email{jugander@fb.com} 
\alignauthor Sebastiano Vigna\thanks{Paolo Boldi, Marco Rosa and Sebastiano
Vigna have been partially supported by a Yahoo!~faculty grant and by MIUR
PRIN ``Query log e web crawling''.}\\
\affaddr{DSI, Universit\`a degli Studi di Milano, Italy}\\
	\email{vigna@acm.org}
}
\else
\author{%
Lars Backstrom\thanks{Facebook.}\quad Paolo Boldi\thanks{DSI, Universit\`a
degli Studi di Milano, Italy. Paolo Boldi, Marco Rosa and Sebastiano
Vigna have been partially supported by a Yahoo!~faculty grant and by MIUR
PRIN ``Query log e web crawling''.} \quad Marco Rosa${}^\dag$ \quad Johan
Ugander${}^*$ \quad Sebastiano Vigna${}^\dag$}
\fi

\begin{document}
\clubpenalty=10000 
\widowpenalty = 10000
\bibliographystyle{plain}
\maketitle

\begin{abstract}
Frigyes Karinthy, in his 1929 short story ``L\'ancszemek'' (``Chains'') suggested
that any two persons are distanced by at most six friendship
links.\footnote{The exact wording of the story is slightly ambiguous: ``He bet
us that, using no more than five individuals, one of whom is a personal
acquaintance, he could contact the selected individual [\ldots]''. It is not
completely clear whether the selected individual is part of the five, so this
could actually allude to distance five or six in the language of graph
theory, but the ``six degrees of separation'' phrase stuck after John Guare's
1990 eponymous play. Following Milgram's definition and Guare's interpretation
(see further on), we will assume that ``degrees of separation'' is the same as ``distance
minus one'', where ``distance'' is the usual path length (the number of arcs
in the path).} Stanley Milgram in his famous experiment~\cite{MilSWP,TMESSWP}
challenged people to route postcards to a fixed recipient by passing them only 
through direct acquaintances. The average number of intermediaries
on the path of the postcards lay between $4.4$ and $5.7$,
depending on the sample of people chosen.

We report the results of the first world-scale social-network graph-distance
computations, using the entire Facebook network of active users ($\approx721$
million users, $\approx 69$ billion friendship links). The average distance we
observe is $4.74$, corresponding to $3.74$ intermediaries or  ``degrees
of separation'', showing that the
world is even smaller than we expected, and prompting the title of this paper.
More generally, we study the distance distribution of Facebook and of some
interesting geographic subgraphs, looking also at their evolution over time.

The networks we are able to explore are almost two orders of magnitude larger
than those analysed in the previous literature. We report detailed statistical
metadata showing that our measurements (which rely on probabilistic
algorithms) are very accurate.
\end{abstract}

%\category{G.2.2}{Graph theory}{Graph algorithms}
%\category{G.3}{Probability and Statistics}{Probabilistic algorithms}
% 
%\terms{Algorithms, Experiments}
%\keywords{Neighbourhood function, probabilistic counters, shortest paths, effective diameter}
% 

\section{Introduction}

At the 20th World--Wide Web Conference, in Hyderabad, India, one of the authors
(Sebastiano) presented a new tool for studying the distance distribution of very
large graphs: HyperANF~\cite{BRVH}. Building on previous graph
compression~\cite{BoVWFI} work and on the idea of diffusive computation pioneered
in~\cite{PGFANF}, the new tool made it possible to accurately study the distance
distribution of graphs orders of magnitude larger than it was previously
possible.

One of the goals in studying the distance distribution is the identification of
interesting statistical parameters that can be used to tell proper social
networks from other complex networks, such as web graphs. More generally,
the distance distribution is one interesting \emph{global} feature that makes it
possible to reject probabilistic models even when they match local features such as the
in-degree distribution.

In particular, earlier work had shown that the \emph{spid}\footnote{The 
spid (shortest-paths index of dispersion) is the variance-to-mean ratio of 
the distance distribution.}, which measures the \emph{dispersion} of the distance 
distribution, appeared to be smaller than 1 (underdispersion) for social networks, 
but larger than one (overdispersion) for web graphs~\cite{BRVH}. Hence, during 
the talk, one of the main open questions was ``What is the spid of Facebook?''.

Lars Backstrom happened to listen to the talk, and suggested a
collaboration studying the Facebook graph. This was of course an extremely
intriguing possibility: beside testing the ``spid hypothesis'', computing the distance
distribution of the Facebook graph would have been the largest Milgram-like~\cite{MilSWP}
experiment ever performed, orders of magnitudes larger than
previous attempts (during our experiments Facebook has $\approx721$ million
active users and $\approx 69$ billion friendship links).

This paper reports our findings in studying the distance distribution of the
largest electronic social network ever created. That world is smaller than we
thought: the average distance of the current Facebook graph is $4.74$. Moreover,
the spid of the graph is just $0.09$, corroborating the conjecture~\cite{BRVH}
that proper social networks have a spid well below one. We also observe,
contrary to previous literature analysing graphs orders of magnitude smaller,
both a stabilisation of the average distance over time, and that the density of
the Facebook graph over time does not neatly fit previous models.

%and a quite erratic change in
%density as the Facebook graph grows.

Towards a deeper understanding of the structure of the Facebook graph, we also
apply recent compression techniques that exploit the underlying cluster
structure of the graph to increase \emph{locality}. The results obtained
suggests the existence of overlapping clusters similar to those observed in other social
networks.

Replicability of scientific results is important. While for obvious
nondisclosure reasons we cannot release to the public the actual 30 graphs
that have been studied in this paper, we distribute freely the derived
data upon which the tables and figures of this papers have been built,
that is, the WebGraph \emph{properties}, which contain structural
information about the graphs, and the probabilistic estimations of their
neighbourhood functions (see below) that have been used to study their
distance distributions. The software used in this paper is distributed
under the (L)GPL General Public License.\footnote{See
\texttt{http://\{webgraph,law\}.dsi.unimi.it/}.}

\section{Related work}
\label{sec:related}

The most obvious precursor of our work is Milgram's celebrated ``small
world'' experiment, described first in~\cite{MilSWP} and later with more details
in~\cite{TMESSWP}: Milgram's works were actually following a stream of research
started in sociology and psychology in the late 50s~\cite{GurSSAN}. In his
experiment, Milgram aimed at answering the following
question (in his words): ``given two individuals selected randomly from the
population, what is the probability that the minimum number of intermediaries
required to link them is 0, 1, 2, \dots, $k$?''.

The technique Milgram used (inspired by~\cite{RHSLS}) was the following: he
selected 296 volunteers (the \emph{starting population}) and asked them to
dispatch a message to a specific individual (the \emph{target person}), a
stockholder living in Sharon, MA, a suburb of Boston, and working in Boston. 
The message could not be sent directly to the target person (unless the sender
knew him personally), but could only be mailed to a personal acquaintance who
is more likely than the sender to know the target person. 
The starting population was selected as follows: 100 of them were people living
in Boston, 100 were Nebraska stockholders (i.e., people living far from
the target but sharing with him their profession) and 96 were Nebraska
inhabitants chosen at random.

In a nutshell, the results obtained from Milgram's experiments were the
following: only 64 chains ($22\%$) were completed (i.e., they reached the
target); the average number of intermediaries in these chains was $5.2$, with a
marked difference between the Boston group ($4.4$) and the rest of the starting
population, whereas the difference between the two other subpopulations was not
statistically significant; at the other end of the spectrum, the random
(and essentially clueless) group from Nebraska needed $5.7$ intermediaries on
average (i.e., rounding up, ``six degrees of separation''). The main conclusions
outlined in Milgram's paper were that the average path length is small, much smaller than expected, and that geographic
location seems to have an impact on the average length whereas other information
(e.g., profession) does not.

There is of course a fundamental difference between our experiment and what
Milgram did: Milgram was measuring the average length of a
\emph{routing path} on a social network, which is of course an upper bound
on the average distance (as the people involved in the experiment were not
necessarily sending the postcard to an acquaintance on a shortest path to the
destination).\footnote{Incidentally, this observation is at the basis of
one of the most intense monologues in Guare's play: Ouisa, unable to locate
Paul, the con man who convinced them he is the son of Sidney Poitier, says
``I read somewhere that everybody on this planet is separated by only six other people. 
Six degrees of separation. Between us and everybody else on this planet.
[\ldots] But to find the right six people.'' Note that this
fragment of the monologue clearly shows that Guare's interpretation of the ``six
degree of separation'' idea is equivalent to distance \emph{seven} in the
graph-theoretical sense.} In a sense, the results he obtained are even
more striking, because not only do they prove that the world is small, 
but that the actors living in the small world are able to exploit its smallness. 
It should be remarked, however, that in~\cite{MilSWP,TMESSWP} the purpose
of the authors is to estimate the number of intermediaries: the postcards are
just a tool, and the details of the paths they follow are studied only as an
artifact of the measurement process. The interest in efficient routing
lies more in the eye of the beholder (e.g., the computer scientist) than in
Milgram's: with at his disposal an actual large 
database of friendship links and algorithms like the ones we use, he would have
dispensed with the postcards altogether.

Incidentally, there have been some attempts to reproduce Milgram-like
routing experiments on various large networks~\cite{LNKRTGRSN,LPSMSN,GMWSSSWE},
but the results in this direction are still very preliminary because notions
such as identity, knowledge or routing are still poorly understood in social
networks.

We limited ourselves to the part of Milgram's experiment that is more clearly
defined, that is, the measurement of shortest paths. The largest experiment
similar to the ones presented here that we are aware of is~\cite{LHPSVLIMN},
where the authors considered a \emph{communication graph} with $180$ million
nodes and $1.3$ billion edges extracted from a snapshot of the Microsoft
Messenger network; they find an average distance of $6.6$ (i.e., $5.6$
intermediaries; again, rounding up, six degrees of separation).
% \footnote{The
% authors make some confusion between ``hops'' and ``degrees of separation'',
% which makes them claim to have found \emph{seven} degrees of separation.}
Note, however, that the communication graph in~\cite{LHPSVLIMN} has an edge
between two persons only if they communicated during a specific one-month
observation period, and thus does not take into account friendship links through
which no communication was detected.

The authors of \cite{YWWDDASPLERWN}, instead, study the distance
distribution of some small-sized social networks. In both cases the networks
were undirected and small enough (by at least two orders of magnitude) to be
accessed efficiently in a random fashion, so the authors used \emph{sampling}
techniques. We remark, however, that sampling is not easily applicable
to  directed networks (such as Twitter) that are not strongly connected, whereas
our techniques would still work (for some details about the applicability
of sampling, see~\cite{CGLCTAADDRWG}).

%I think it's enough that our data is bigger, don't need to point out flaws, in case authors are reviewers --Lars

%It must be remarked, however, that in both cases the authors do not
%eport necessary statistical data, such as confidence intervals or standard
%eviation, from which the precision of the measurement can be assessed and
%ompared with ours. 

Analysing the evolution of social networks in time is also a lively trend of
research. Leskovec, Kleinberg and Faloutsos observe in~\cite{LKFGEDSD} that the
average degree of complex networks increase over time while the
\emph{effective diameter} shrinks. Their experiments are conducted  on a much
smaller scale (their largest graph has $4$ millions of nodes and $16$ millions of arcs), but it is
interesting that the phenomena observed seems quite consistent. Probably the
most controversial point is the hypothesis that the number of edges $m(t)$ at
time $t$ is related to the number of nodes $n(t)$ by the following relation: \[
 m(t) \propto n(t)^a, \] where $a$ is a fixed exponent usually lying in
 the interval $(1\,.\,.\,2)$. We will discuss this hypothesis in light of our
 findings.
 
\section{Definitions and Tools}

The \emph{neighbourhood function} $N_G(t)$ of a graph $G$ returns for each
$t\in\mathbf N$ the number of pairs of nodes $\langle x, y\rangle$ such that
$y$ is reachable from $x$ in at most $t$ steps. It provides data about
how fast the ``average ball'' around each node expands. From the neighbourhood
function it is possible to derive the distance distribution (between reachable
pairs), which gives for each $t$ the fraction of reachable pairs at distance
exactly $t$.

In this paper we use HyperANF, a diffusion-based algorithm (building on
ANF~\cite{PGFANF}) that is able to approximate quickly the neighbourhood
function of very large graphs; our implementation uses, in turn,
WebGraph~\cite{BoVWFI} to represent in a compressed but quickly accessible form
the graphs to be analysed.

HyperANF is based on the observation (made in~\cite{PGFANF}) that $B(x,r)$, the ball
of radius $r$ around node $x$, satisfies
\[
B(x,r) = \bigcup_{x\to y}B(y,r-1) \cup \{\,x\,\}.
\]
Since $B(x,0)=\{\,x\,\}$, we can compute each $B(x,r)$ incrementally using
sequential scans of the graph (i.e., scans in which we go in turn through the
successor list of each node). The obvious problem is that during the scan we
need to access randomly the sets $B(x,r-1)$ (the sets $B(x,r)$ can be just
saved on disk on a \emph{update file} and reloaded later). 

The space needed for such sets would be too large to be kept in main memory.
However, HyperANF represents these sets in an \emph{approximate} way, using
\emph{HyperLogLog counters}~\cite{FFGH}, which should be thought as dictionaries
that can answer reliably just questions about size. Each such counter is made
of a number of small (in our case, 5-bit) \emph{registers}. In a
nutshell, a register keeps track of the maximum number $M$ of trailing zeroes of the values of a good hash
function applied to the elements of a sequence of nodes: the number of distinct
elements in the sequence is then proportional to $2^M$. A technique called
\emph{stochastic averaging} is used to divide the stream into a number of
substreams, each analysed by a different register. The result is then computed
by aggregating suitably the estimation from each register (see~\cite{FFGH} for
details).

The main performance challenge to solve is how to quickly compute the HyperLogLog
counter associated to a union of balls, each represented, in turn, by a
HyperLogLog counter: HyperANF uses an algorithm based on word-level parallelism
that makes the computation very fast, and a carefully engineered implementation
exploits multicore architectures with a linear speedup in the number of cores.

Another important feature of HyperANF is that it uses a \emph{systolic} approach
to avoid recomputing balls that do not change during an iteration. This approach
is fundamental to be able to compute the entire distance distribution, avoiding
the arbitrary termination conditions used by previous approaches, which have no
provable accuracy (see~\cite{BRVH} for an example).

\subsection{Theoretical error bounds}

The result of a run of HyperANF at the $t$-th iteration is an estimation of the
neighbourhood function in $t$. We can see it as a random variable 
\[
\Nhat_G(t)=\sum_{0\leq i< n} X_{i,t}
\] 
where each
$X_{i,t}$ is the HyperLogLog counter that counts nodes reached by node $i$ in
$t$ steps ($n$ is the number of nodes of the graph). When $m$ registers per
counter are used, each $X_{i,t}$ has a guaranteed relative standard deviation $\eta_m\leq 1.06/\sqrt m$.

It is shown in~\cite{BRVH} that the output $\Nhat_G(t)$ of HyperANF at the
$t$-th iteration is an asymptotically almost unbiased estimator of $N_G(t)$, that is
\[
	\frac{E[\Nhat_G(t)]}{N_G(t)}=1+\delta_1(n)+o(1) \text{ for $n\to \infty$},
\]
where $\delta_1$ is the same as in~\cite{FFGH}[Theorem 1] (and
$|\delta_1(x)|<5\cdot 10^{-5}$ as soon as $m\geq 16$). Moreover, $\Nhat_G(t)$
has a relative standard deviation not greater than that of the $X_i$'s, that
is
\[\frac{\StdDev{\Nhat_G(t)}}{N_G(t)} \leq \eta_m.\]

In particular, our runs used $m=64$ ($\eta_m=0.1325$) for all graphs except for
the two largest Facebook graphs, where we used $m=32$ ($\eta_m=0.187$).
Runs were repeated so to obtain a uniform relative standard deviation for all
graphs.

% The error
% passes unimodality tests, so we can apply
% the Vysochanski{\u\i}-Petunin inequality~\cite{VPJTSRUD}, obtaining the bound
% \[
% \Pr\left[\frac{\Nhat_G(t)}{N_G(t)} \in (1-\epsilon,1+\epsilon)\right] \geq
% 1-\frac{4\eta_m^2}{9\epsilon^2} .
% \]
Unfortunately, the relative error for the neighbourhood function becomes an
\emph{absolute} error for the distance distribution. Thus, the theoretical
bounds one obtains for the moments of the distance distribution are quite ugly.
Actually, the simple act of dividing the neighbourhood function values by the
last value to obtain the cumulative distribution function is nonlinear, and
introduces bias in the estimation.

To reduce bias and provide estimates of the standard error of our measurements,
we use the \emph{jackknife}~\cite{EfGLLBJCV}, a classical nonparametric method
for evaluating arbitrary statistics on a data sample, which turns out to be very
effective in practice~\cite{BRVH}.

\section{Experiments}
\label{sec:experiments}

The graphs analysed in this paper are graphs of Facebook users who were
active in May of 2011; an active user is one who 
has logged in within the last 28 days. The decision to restrict our
study to active users allows us to eliminate accounts that have been
abandoned in early stages of creation, and focus on accounts that
plausibly represent actual individuals. In accordance with Facebook's data
retention policies, historical user activity records are not retained, and
historical graphs for each year were constructed by considering currently
active users that were registered on January 1st of that year, along with
those friendship edges that were formed prior that that date. The
``current'' graph is simply the graph of active users at the time when the
experiments were performed (May 2011). The graph predates the existence of
Facebook ``subscriptions'', a directed relationship feature introduced in
August 2011, and also does not include ``pages'' (such as celebrities)
that people may ``like''. For standard user accounts on Facebook there is
a limit of 5\,000 possible friends.

We decided to extend our experiments in two directions: regional and temporal.
We thus analyse the entire Facebook graph (\texttt{fb}), the USA subgraph
(\texttt{us}), the Italian subgraph (\texttt{it}) and the Swedish (\texttt{se})
subgraph. We also analysed a combination of the Italian and Swedish graph
(\texttt{itse}) to check whether combining two regional but distant networks could
significantly change the average distance, in the same spirit as in the original
Milgram's experiment.\footnote{To establish geographic location, we use the
users' \emph{current} geo-IP location; this means, for example, that the users
in the it-2007 graph are users who are today in Italy and were on Facebook on
January 1, 2007 (most probably, American college students then living in
Italy).} For each graph we compute the distance distribution from 2007 up to
today by performing several HyperANF runs, obtaining an estimate of values of
neighbourhood function with relative standard deviation at most $5.8\%$: in
several cases, however, we performed more runs, obtaining a higher precision. We
report the jackknife~\cite{EfGLLBJCV} estimate of derived values (such as
average distances) and the associated estimation of the standard error.

% SEBA: tell more

\subsection{Setup}

The computations were performed on a 24-core machine with 72\,GiB of memory and
1\,TiB of disk space.\footnote{We remark that the commercial value of such
hardware is of the order of a few thousand dollars.} The first task was to
import the Facebook graph(s) into a compressed form for WebGraph~\cite{BoVWFI},
so that the multiple scans required by HyperANF's diffusive process could be
carried out relatively quickly. This part required some massaging of Facebook's
internal IDs into a contiguous numbering: the resulting current \texttt{fb}
graph (the largest we analysed) was compressed to $345$\,GB at 20 bits per arc,
which is $86\%$ of the information-the\-o\-ret\-i\-cal lower bound
($\log{{n^2} \choose m}$ bits, there $n$ is the number of nodes and
$m$ the number of arcs).\footnote{Note that we measure compression with respect to the lower bound
on \emph{arcs}, as WebGraph stores \emph{directed} graphs; however, with the
additional knowledge that the graph is undirected, the lower bound should be
applied to \emph{edges}, thus doubling, in practice, the number of bits used.}
Whichever coding we choose, for half of the possible
graphs with $n$ nodes and $m$ arcs we need at least $\bigl\lfloor\log{{n^2}
\choose m}\bigr\rfloor$ bits per graph: the purpose of
compression is precisely to choose the coding so to represent interesting graphs
in a smaller space than that required by the bound.

To understand what is happening, we recall that WebGraph uses the BV compression
scheme~\cite{BoVWFI}, which applies three intertwined techniques to the
successor list of a node:
\begin{itemize}
  \item successors are (partially) \emph{copied} from previous nodes within a
  small window, if successors lists are similar enough;
  \item successors are \emph{intervalised}, that is, represented by a left
  extreme and a length, if significant contiguous successor sequences appear;
  \item successors are \emph{gap-compressed} if they pass
  the previous phases: instead of storing the actual successor list, we store
  the differences of consecutive successors (in increasing order) using
  instantaneous codes.
\end{itemize}
Thus, a graph compresses well when it exhibits \emph{similarity} (nodes with
near indices have similar successor lists) and \emph{locality} (successor
lists have small gaps).

The better-than-random result above (usually, randomly permuted graphs
compressed with WebGraph occupy $10-20\%$ more space than the lower bound) has most likely been
induced by the renumbering process, as in the original stream of arcs 
all arcs going out from a node appeared consecutively; as a
consequence, the renumbering process assigned consecutive labels to all
yet-unseen successors (e.g., in the initial stages successors were labelled
contiguously), inducing some locality.

It is also possible that the ``natural'' order for Facebook (essentially, join
order) gives rise to some improvement over the information-theoretical lower
bound because users often join the network at around the same time as several
of their friends, which causes a certain amount of locality and similarity, as
circle of friends have several friends in common.

We were interested in the first place to establish whether more locality could
be induced by suitably permuting the graph using \emph{layered labelled
propagation}~\cite{BRSLLP} (LLP). This approach (which computes several
clusterings with different levels of granularity and combines them to sort the
nodes of a graph so to increase its locality and similarity) has recently led to
the best compression ratios for social networks when combined with the BV compression
scheme. An increase in compression means that we were able to partly understand
the cluster structure of the graph.

We remark that each of the clusterings required by LLP is in itself a
\textit{tour de force}, as the graphs we analyse are almost two orders of
magnitude larger than any network used for experiments in the literature on
graph clustering. Indeed, applying LLP to the current Facebook graph
required ten days of computation on our hardware.

\begin{table*}[ht]
\begin{center}
\begin{tabular}{|c|r|r|r|r|r|}
	\hline
	& \multicolumn{1}{c|}{\texttt{it}} & \multicolumn{1}{c|}{\texttt{se}} &
	\multicolumn{1}{c|}{\texttt{itse}} & \multicolumn{1}{c|}{\texttt{us}}&
	\multicolumn{1}{c|}{\texttt{fb}} \\
	\hline
	Original & 14.8 ($83\%$) & 14.0 ($86\%$) & 15.0 ($82\%$) & 17.2
	($82\%$) & 20.1 ($86\%$) \\ LLP         & 10.3 ($58\%$) & 10.2 ($63\%$) & 10.3
	($56\%$) & 11.6 ($56\%$)  & 12.3 ($53\%$)\\
	\hline
\end{tabular}
\end{center}
\caption{\label{tab:comp}The number of bits per link and the  
compression ratio (with respect to the information-theoretical lower bound) for
the current graphs in the original order and for the same
graphs permuted by layered label propagation~\cite{BRSLLP}.}
\end{table*}

We applied layered labelled propagation and re-compressed our graphs (the current
version), obtaining a significant improvement. In
Table~\ref{tab:comp} we show the results: we were able to reduce the graph size by $30\%$, which
suggests that LLP has been able to discover several significant clusters.
 
\begin{figure}
\centering
\includegraphics[scale=.5]{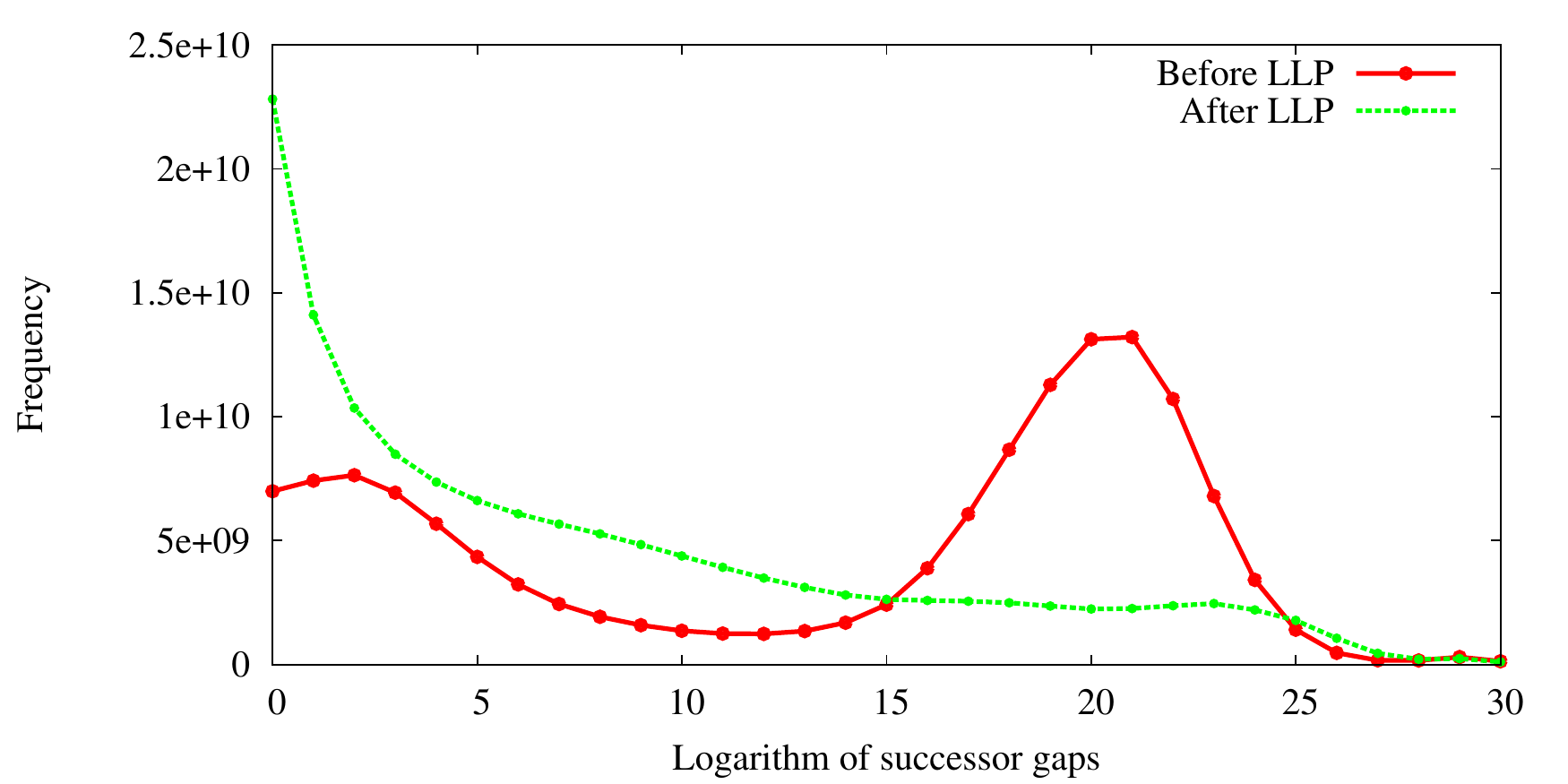}
\caption{\label{fig:gap} The change in distribution of the logarithm of
the gaps between successors when the current \texttt{fb} graph is permuted by
layered label propagation. See also Table~\ref{tab:comp}.}
\end{figure}

The change in structure can be easily seen from Figure~\ref{fig:gap}, where we
show the distribution of the binary logarithm of gaps between successors for
the current \texttt{fb} graph. The
smaller the gaps, the higher the locality. In the graph with renumbered Facebook IDs, the distribution is bimodal: there is a 
local maximum at two, showing that there is some locality, but the bulk of the
probability mass is around 20--21, which is slightly less than the
information-theoretical lower bound ($\approx 23$).

In the graph permuted with LLP, however, the distribution radically changes: it
is now (mostly) beautifully monotonically decreasing, with a very small bump at
23, which testifies the existence of a small core of ``randomness'' in the graph
that LLP was not able to tame.

Regarding similarity, we see an analogous phenomenon: the number of
successors represented by copy has doubled, going from
$9\%$ to $18\%$. The last datum is in line
with other social networks (web graphs, on the contrary, are extremely redundant
and more than $80\%$ of the successors are usually copied). Moreover, disabling
copying altogether results in modest increase in size ($\approx 5\%$), again in
line with other social networks, which suggests that for most applications it is better to
disable copying at all to obtain faster random access.

The compression ratio is around $53\%$, which is similar to
other similar social networks, such as LiveJournal ($55\%$) or DBLP
($40\%$)~\cite{BRSLLP}\footnote{The interested reader will find similar data
for several type of networks at the LAW web site
(\texttt{http://law.dsi.unimi.it/}).}. For other graphs (see
Table~\ref{tab:comp}), however, it is slightly worse. This might be due to
several phenomena: First, our LLP runs were executed with only half the number or
clusters, and for each cluster we restricted the number of iterations to just four, 
to make the whole execution of LLP feasible. Thus, our runs are capable of
finding considerably less structure than the runs we had previously performed
for other networks. Second, the number of nodes is much larger: there is some cost in
writing down gaps (e.g., using $\gamma$, $\delta$ or $\zeta$ codes) that is
dependent on their absolute magnitude, and the lower bound does not take
into account that cost.

\subsection{Running}

% In the version presented at WWW 2011, HyperANF was not able to handle Facebook's
% graph on the available hardware. The reason is that, for programming simplicity,
% we chose to support no less than 64 registers per counter (this makes for much
% simpler code, as independently of the register size counters end up being 64-bit
% aligned). Due to the hardware available, however, we had to modify HyperANF so
% to make it possible to use as little as 16 registers (at the expenses of
% relative standard deviation, of course). 

Since most of the graphs, because of their size, had to be accessed by memory
mapping, we decided to store all counters (both those for $B(x,r-1)$ and those
for $B(x,r)$) in main memory, to avoid eccessive I/O. The runs of HyperANF on
the current whole Facebook graph used 32 registers, so the space for counters was about
$27$\,GiB (e.g., we could have analysed a graph with four times the number of
nodes on the same hardware). As a rough measure of speed, a run on the LLP-compressed
current whole Facebook graph requires about $13.5$ hours. Note that this timings would scale linearly with an increase in
the number of cores.

% We quickly noticed is that such an unusual (for a
% Java application) memory footprint was causing troubles: the garbage collector
% needs to allocate memory for the ``young'' generation, that is, objects that
% will be quickly created and quickly collected. Unfortunately, the heuristics to
% decide the size of the young generation is that of sizing it proportionally to
% the ``old generation'', which in the case of HyperANF is dozens of gigabytes.
% Fortunately, the Sun JVM makes it possible to manually size the young generation
% (e.g., \texttt{-XX:MaxNewSize=1G}), which was enough to significantly reduce the
% memory consumption of the JVM.

\begin{figure}[tb]
\centering
\includegraphics[scale=.45]{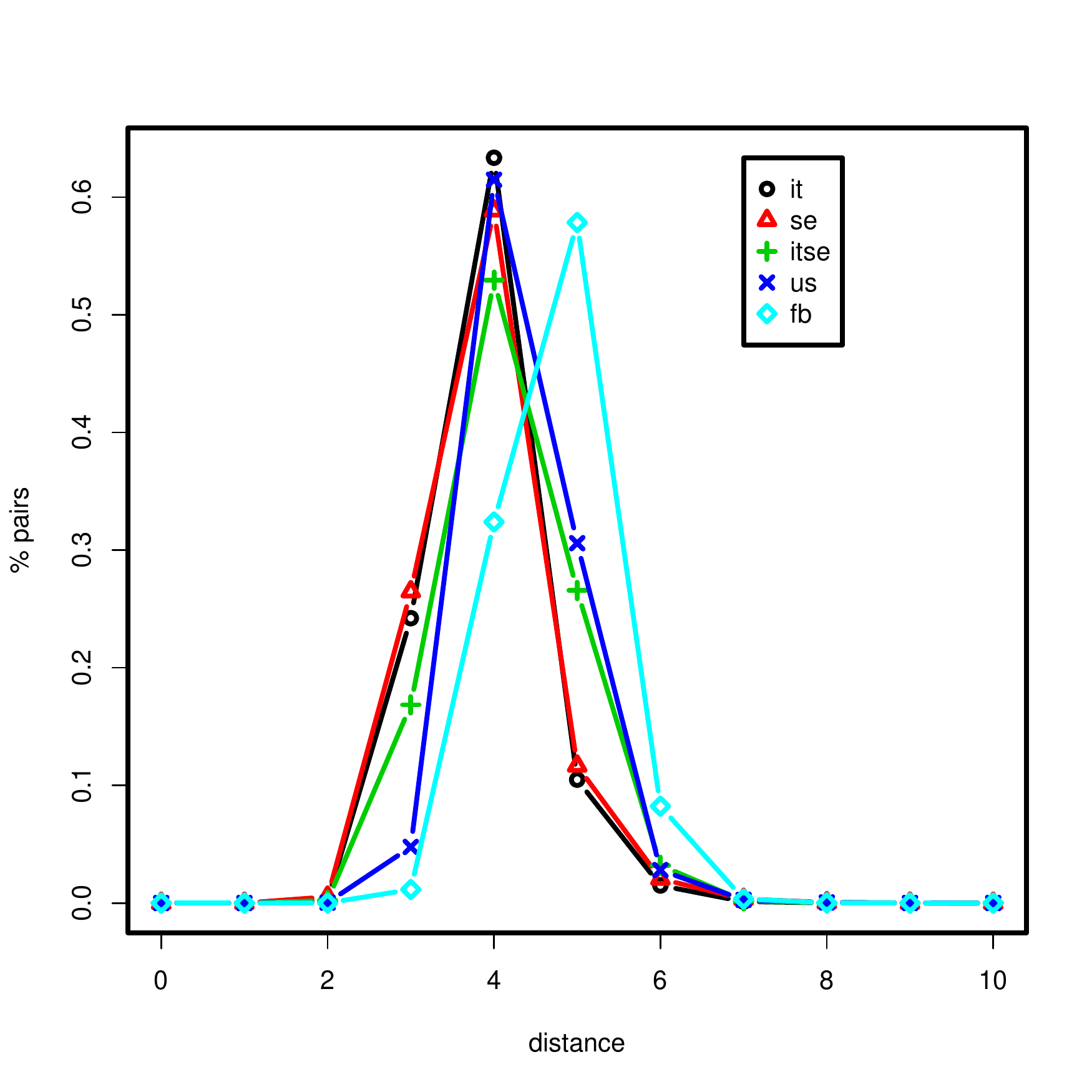}
\caption{\label{fig:pdf} The probability mass functions of the distance
distributions of the current graphs (truncated at distance 10).}
\end{figure}

\begin{figure}[tb]
\centering
\includegraphics[scale=.45]{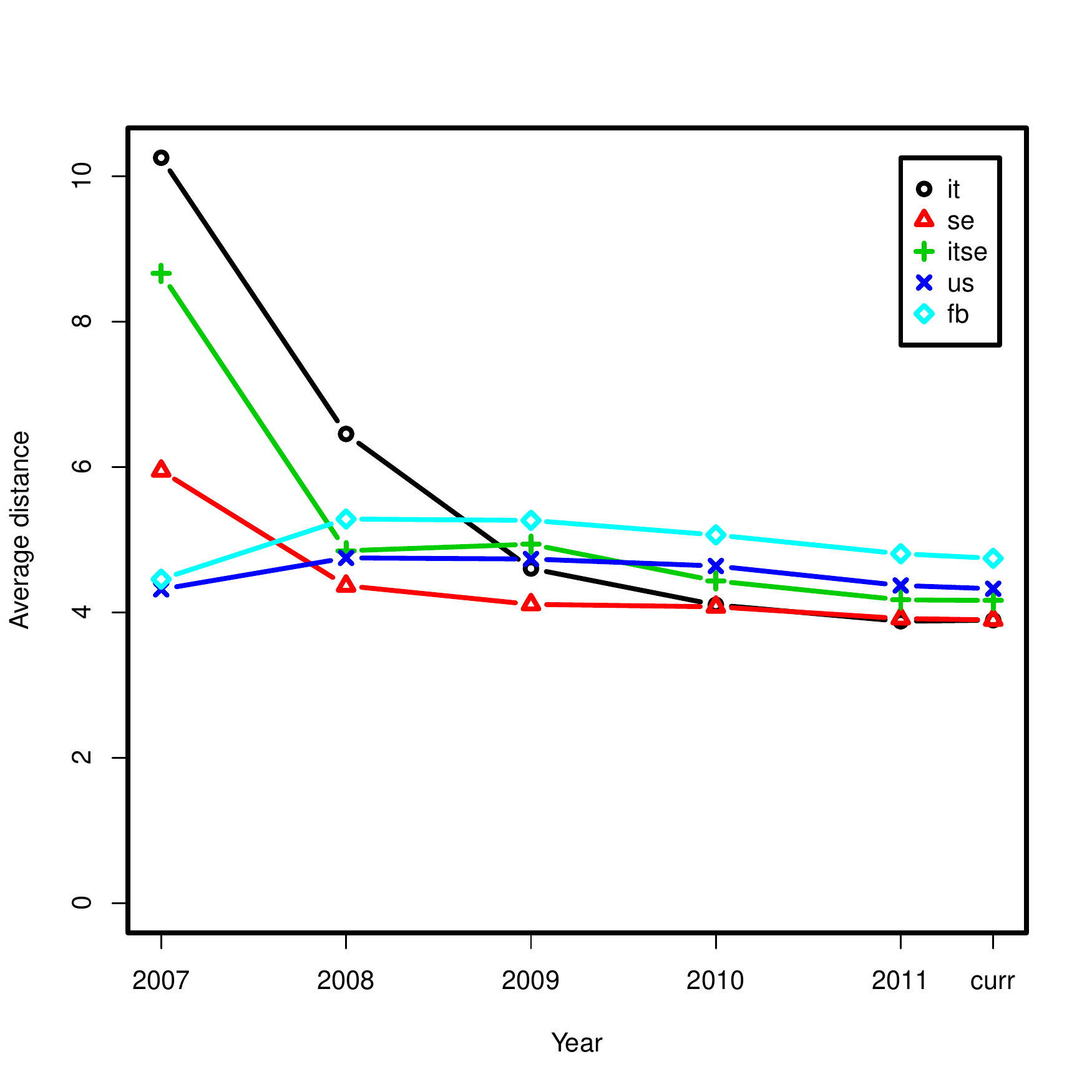}
\caption{\label{fig:avgdist} The average distance graph. See also
Table~\ref{tab:avg}.}
\end{figure}

\begin{figure}[tb]
\centering
\includegraphics[scale=.45]{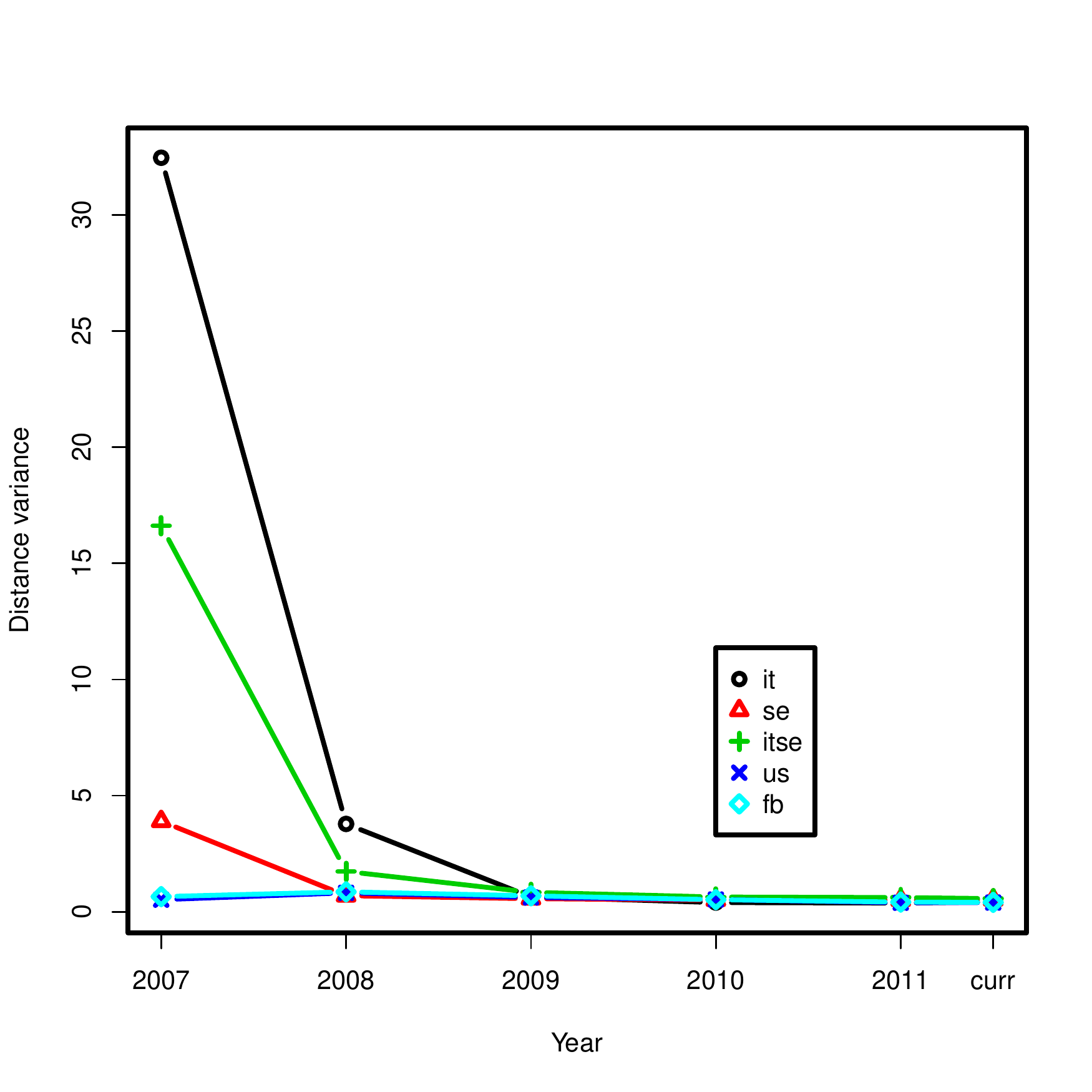}
\caption{\label{fig:var}The graph of variances of the distance
distributions. See also Table~\ref{tab:var}.}
\end{figure}

\subsection{General comments}

In September 2006, Facebook was opened to non-college students: there was an
instant surge in subscriptions, as our data shows. In particular, the 
\texttt{it} and \texttt{se} subgraphs from January 1, 2007 were highly disconnected, 
as shown by the incredibly low percentage of reachable pairs we estimate in
Table~\ref{tab:reachable}. Even Facebook itself was rather disconnected, but all the data
we compute stabilizes (with small oscillations) after 2009, with essentially
all pairs reachable. Thus, we consider the data for 2007 and
2008 useful to observe the evolution of Facebook, but we do not consider them representative of the underlying human social-link structure.

\begin{table*}[ht]
\begin{center}
\begin{tabular}{|c|r|r|r|r|r|}
	\hline
	& \multicolumn{1}{c|}{\texttt{it}} & \multicolumn{1}{c|}{\texttt{se}} & \multicolumn{1}{c|}{\texttt{itse}} & \multicolumn{1}{c|}{\texttt{us}} & \multicolumn{1}{c|}{\texttt{fb}} \\
	\hline
	2007 & 159.8\,K (105.0\,K) & 11.2\,K (21.8\,K) & 172.1\,K (128.8\,K) & 8.8\,M (529.3\,M) & 13.0\,M (644.6\,M) \\
2008 & 335.8\,K (987.9\,K) & 1.0\,M (23.2\,M) & 1.4\,M (24.3\,M) & 20.1\,M (1.1\,G) & 56.0\,M (2.1\,G) \\
2009 & 4.6\,M (116.0\,M) & 1.6\,M (55.5\,M) & 6.2\,M (172.1\,M) & 41.5\,M (2.3\,G) & 139.1\,M (6.2\,G) \\
2010 & 11.8\,M (726.9\,M) & 3.0\,M (149.9\,M) & 14.8\,M (878.4\,M) & 92.4\,M (6.0\,G) & 332.3\,M (18.8\,G) \\
2011 & 17.1\,M (1.7\,G) & 4.0\,M (278.2\,M) & 21.1\,M (2.0\,G) & 131.4\,M (12.4\,G) & 562.4\,M (47.5\,G) \\
current & 19.8\,M (2.2\,G) & 4.3\,M (335.7\,M) & 24.1\,M (2.6\,G) & 149.1\,M (15.9\,G) & 721.1\,M (68.7\,G) \\

	\hline
\end{tabular}
\end{center}
\caption{\label{tab:nodeslinks} Number of nodes and friendship links of the
datasets. Note that each friendship link, being undirected, is represented by a pair of
symmetric arcs.}
\end{table*}

\begin{table*}[ht]
\begin{center}
\begin{tabular}{|c|r|r|r|r|r|}
	\hline
	& \multicolumn{1}{c|}{\texttt{it}} & \multicolumn{1}{c|}{\texttt{se}} & \multicolumn{1}{c|}{\texttt{itse}} & \multicolumn{1}{c|}{\texttt{us}} & \multicolumn{1}{c|}{\texttt{fb}} \\
	\hline
	2007 & 387.0\,K & 51.0\,K & 461.9\,K & 1.8\,G & 2.3\,G \\
2008 & 3.9\,M & 96.7\,M & 107.8\,M & 4.0\,G & 9.2\,G \\
2009 & 477.9\,M & 227.5\,M & 840.3\,M & 9.1\,G & 28.7\,G \\
2010 & 3.6\,G & 623.0\,M & 4.5\,G & 26.0\,G & 93.3\,G \\
2011 & 8.0\,G & 1.1\,G & 9.6\,G & 53.6\,G & 238.1\,G \\
current & 8.3\,G & 1.2\,G & 9.7\,G & 68.5\,G & 344.9\,G \\

	\hline
\end{tabular}
\end{center}
\caption{\label{tab:size} Size in bytes of the datasets.
%  The percentage shown
% is the ratio between the actual size and the information-theoretical lower
% bound.
}
\end{table*}

\begin{table}[ht]
\begin{center}
\begin{tabular}{|c|r|r|r|r|r|}
	\hline
	& \multicolumn{1}{c|}{\texttt{it}} & \multicolumn{1}{c|}{\texttt{se}} & \multicolumn{1}{c|}{\texttt{itse}} & \multicolumn{1}{c|}{\texttt{us}} & \multicolumn{1}{c|}{\texttt{fb}} \\
	\hline
	2007 & 1.31 & 3.90 & 1.50 & 119.61 & 99.50 \\
2008 & 5.88 & 46.09 & 36.00 & 106.05 & 76.15 \\
2009 & 50.82 & 69.60 & 55.91 & 111.78 & 88.68 \\
2010 & 122.92 & 100.85 & 118.54 & 128.95 & 113.00 \\
2011 & 198.20 & 140.55 & 187.48 & 188.30 & 169.03 \\
current & 226.03 & 154.54 & 213.30 & 213.76 & 190.44 \\

	\hline
\end{tabular}
\end{center}
\caption{\label{tab:degree} Average degree of the datasets.}
\end{table}

\begin{table*}[ht]
\begin{center}
\begin{tabular}{|c|r|r|r|r|r|}
	\hline
	& \multicolumn{1}{c|}{\texttt{it}} & \multicolumn{1}{c|}{\texttt{se}} & \multicolumn{1}{c|}{\texttt{itse}} & \multicolumn{1}{c|}{\texttt{us}} & \multicolumn{1}{c|}{\texttt{fb}} \\
	\hline
	2007&8.224E-06&3.496E-04&8.692E-06&1.352E-05&7.679E-06\\
2008&1.752E-05&4.586E-05&2.666E-05&5.268E-06&1.359E-06\\
2009&1.113E-05&4.362E-05&9.079E-06&2.691E-06&6.377E-07\\
2010&1.039E-05&3.392E-05&7.998E-06&1.395E-06&3.400E-07\\
2011&1.157E-05&3.551E-05&8.882E-06&1.433E-06&3.006E-07\\
current&1.143E-05&3.557E-05&8.834E-06&1.434E-06&2.641E-07\\

	\hline
\end{tabular}
\end{center}
\caption{\label{tab:density}Density of the datasets.}
\end{table*}

\begin{table*}[ht]
\begin{center}
\begin{tabular}{|c|r|r|r|r|r|}
	\hline
	& \multicolumn{1}{c|}{\texttt{it}} & \multicolumn{1}{c|}{\texttt{se}} & \multicolumn{1}{c|}{\texttt{itse}} & \multicolumn{1}{c|}{\texttt{us}} & \multicolumn{1}{c|}{\texttt{fb}} \\
	\hline
			 2007  &  10.25 ($\pm$0.17) &  5.95 ($\pm$0.07) &  8.66 ($\pm$0.14) &  4.32 ($\pm$0.02) &  4.46 ($\pm$0.04) \\
		 2008  &  6.45 ($\pm$0.03) &  4.37 ($\pm$0.03) &  4.85 ($\pm$0.05) &  4.75 ($\pm$0.02) &  5.28 ($\pm$0.03) \\
		 2009  &  4.60 ($\pm$0.02) &  4.11 ($\pm$0.01) &  4.94 ($\pm$0.02) &  4.73 ($\pm$0.02) &  5.26 ($\pm$0.03) \\
		 2010  &  4.10 ($\pm$0.02) &  4.08 ($\pm$0.02) &  4.43 ($\pm$0.03) &  4.64 ($\pm$0.02) &  5.06 ($\pm$0.01) \\
		 2011  &  3.88 ($\pm$0.01) &  3.91 ($\pm$0.01) &  4.17 ($\pm$0.02) &  4.37 ($\pm$0.01) &  4.81 ($\pm$0.04) \\
		 current  &  3.89 ($\pm$0.02) &  3.90 ($\pm$0.04) &  4.16 ($\pm$0.01) &  4.32 ($\pm$0.01) &  4.74 ($\pm$0.02) \\

	\hline
\end{tabular}
\end{center}
\caption{\label{tab:avg} The average distance
($\pm$ standard error). See also Figure~\ref{fig:avgdist}
and~\ref{fig:avgspid}.}
\end{table*}

\begin{table*}[ht]
\begin{center}
\begin{tabular}{|c|r|r|r|r|r|}
	\hline
	& \multicolumn{1}{c|}{\texttt{it}} & \multicolumn{1}{c|}{\texttt{se}} & \multicolumn{1}{c|}{\texttt{itse}} & \multicolumn{1}{c|}{\texttt{us}} & \multicolumn{1}{c|}{\texttt{fb}} \\
	\hline
			 2007  &  32.46 ($\pm$1.49) &  3.90 ($\pm$0.12) &  16.62 ($\pm$0.87) &  0.52 ($\pm$0.01) &  0.65 ($\pm$0.02) \\
		 2008  &  3.78 ($\pm$0.18) &  0.69 ($\pm$0.04) &  1.74 ($\pm$0.15) &  0.82 ($\pm$0.02) &  0.86 ($\pm$0.03) \\
		 2009  &  0.64 ($\pm$0.04) &  0.56 ($\pm$0.02) &  0.84 ($\pm$0.02) &  0.62 ($\pm$0.02) &  0.69 ($\pm$0.05) \\
		 2010  &  0.40 ($\pm$0.01) &  0.50 ($\pm$0.02) &  0.64 ($\pm$0.03) &  0.53 ($\pm$0.02) &  0.52 ($\pm$0.01) \\
		 2011  &  0.38 ($\pm$0.03) &  0.50 ($\pm$0.02) &  0.61 ($\pm$0.02) &  0.39 ($\pm$0.01) &  0.42 ($\pm$0.03) \\
		 current  &  0.42 ($\pm$0.03) &  0.52 ($\pm$0.04) &  0.57 ($\pm$0.01) &  0.40 ($\pm$0.01) &  0.41 ($\pm$0.01) \\

	\hline
\end{tabular}
\end{center}
\caption{\label{tab:var} The variance of the distance distribution
($\pm$ standard error). See also Figure~\ref{fig:var}.}
\end{table*}

\begin{table*}[ht]
\begin{center}
\begin{tabular}{|c|r|r|r|r|r|}
	\hline
	& \multicolumn{1}{c|}{\texttt{it}} & \multicolumn{1}{c|}{\texttt{se}} & \multicolumn{1}{c|}{\texttt{itse}} & \multicolumn{1}{c|}{\texttt{us}} & \multicolumn{1}{c|}{\texttt{fb}} \\
	\hline
			 2007  &  3.17 ($\pm$0.106) &  0.66 ($\pm$0.016) &  1.92 ($\pm$0.078) &  0.12 ($\pm$0.003) &  0.15 ($\pm$0.004) \\
		 2008  &  0.59 ($\pm$0.026) &  0.16 ($\pm$0.008) &  0.36 ($\pm$0.028) &  0.17 ($\pm$0.003) &  0.16 ($\pm$0.005) \\
		 2009  &  0.14 ($\pm$0.007) &  0.14 ($\pm$0.004) &  0.17 ($\pm$0.004) &  0.13 ($\pm$0.003) &  0.13 ($\pm$0.009) \\
		 2010  &  0.10 ($\pm$0.003) &  0.12 ($\pm$0.005) &  0.14 ($\pm$0.006) &  0.11 ($\pm$0.004) &  0.10 ($\pm$0.002) \\
		 2011  &  0.10 ($\pm$0.006) &  0.13 ($\pm$0.006) &  0.15 ($\pm$0.004) &  0.09 ($\pm$0.003) &  0.09 ($\pm$0.005) \\
		 current  &  0.11 ($\pm$0.007) &  0.13 ($\pm$0.010) &  0.14 ($\pm$0.003) &  0.09 ($\pm$0.003) &  0.09 ($\pm$0.003) \\

	\hline
\end{tabular}
\end{center}
\caption{\label{tab:spid} The index of dispersion of distances, a.k.a.~spid
($\pm$ standard error). See also Figure~\ref{fig:avgspid}. }
\end{table*}

\begin{table}[ht]
\begin{center}
\begin{tabular}{|c|r|r|r|r|r|}
	\hline
	& \multicolumn{1}{c|}{\texttt{it}} & \multicolumn{1}{c|}{\texttt{se}} & \multicolumn{1}{c|}{\texttt{itse}} & \multicolumn{1}{c|}{\texttt{us}} & \multicolumn{1}{c|}{\texttt{fb}} \\
	\hline
		 2007  &  0.04&  10.23&  0.19&  100.00&  68.02\\
		 2008  &  25.54&  93.90&  80.21&  99.26&  89.04\\
% current & 99.6 & 99.4 & 99.7 & 99.8 & 99.8 \\
	\hline
\end{tabular}
\end{center}
\caption{\label{tab:reachable} Percentage of reachable pairs 2007--2008.}
\end{table}

% \begin{table*}[ht]
% \begin{center}
% \begin{tabular}{|c|r|r|r|r|r|}
% 	\hline
% 	& \multicolumn{1}{c|}{\texttt{it}} & \multicolumn{1}{c|}{\texttt{se}} & \multicolumn{1}{c|}{\texttt{itse}} & \multicolumn{1}{c|}{\texttt{us}} & \multicolumn{1}{c|}{\texttt{fb}} \\
% 	\hline
% 	\input scripts/harmonicno2007.table
% 	\hline
% \end{tabular}
% \end{center}
% \caption{\label{tab:harmonicno2007} Harmonic diameter (2007 omitted).}
% \end{table*}

\begin{table}[ht]
\begin{center}
\begin{tabular}{|c|r|r|r|r|r|}
	\multicolumn{6}{c}{Lower bounds from HyperANF runs}\\
	\hline
	& \multicolumn{1}{c|}{\texttt{it}} & \multicolumn{1}{c|}{\texttt{se}} & \multicolumn{1}{c|}{\texttt{itse}} & \multicolumn{1}{c|}{\texttt{us}} & \multicolumn{1}{c|}{\texttt{fb}} \\
	\hline
			 2007  &  41&  17&  41&  13&  14\\
		 2008  &  28&  17&  24&  17&  16\\
		 2009  &  21&  16&  17&  16&  15\\
		 2010  &  18&  19&  19&  19&  15\\
		 2011  &  17&  20&  17&  18&  35\\
		 current  &  19&  19&  19&  20&  58\\

	\hline
	\multicolumn{6}{c}{Exact diameter of the giant component}\\
	\hline
	current & 25 & 23 & 27 & 30 & 41 \\
	\hline
\end{tabular}
\end{center}
\caption{\label{tab:diameter} Lower bounds for the diameter of all graphs, and
exact values for the giant component ($>99.7\%$) of current graphs computed
using the iFUB algorithm.}
\end{table}

\subsection{The distribution}

Figure~\ref{fig:pdf} displays the probability mass functions of the current
graphs. We will discuss later the variation of the average distance and spid,
but qualitatively we can immediately distinguish the \emph{regional} graphs,
concentrated around distance four, and the \emph{whole} Facebook graph,
concentrated around distance five. The distributions of \texttt{it} and
\texttt{se}, moreover, have significantly less probability mass concentrated on
distance five than \texttt{itse} and \texttt{us}.
The variance data (Table~\ref{tab:var} and Figure~\ref{fig:var}) show that the
distribution became quickly extremely concentrated.

\subsection{Average degree and density}

\begin{figure}
\centering
\includegraphics[scale=.7]{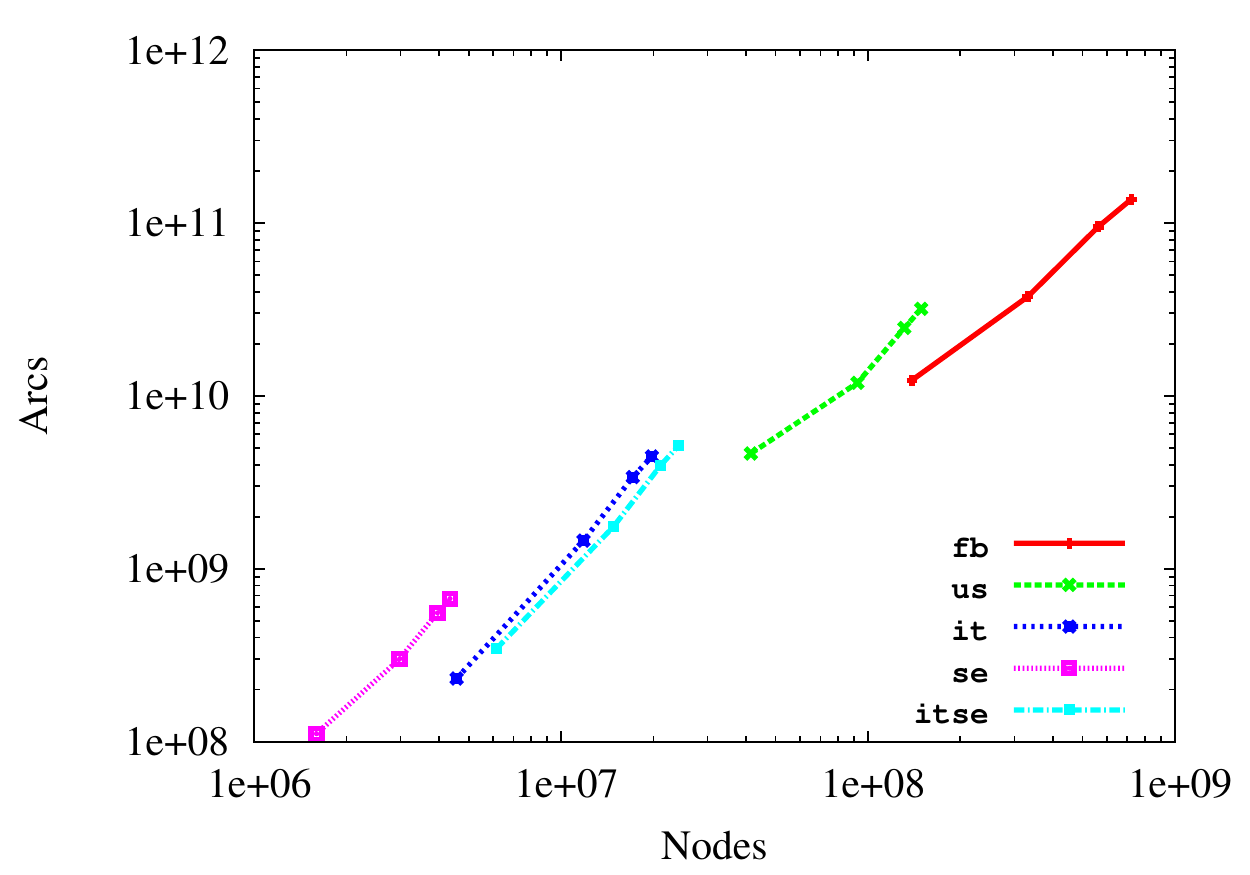}
\caption{\label{fig:degree} A plot correlating number of nodes to the average
degree (for the graphs from 2009 on).}
\end{figure}

Table~\ref{tab:degree} shows the relatively quick growth in time of the average
degree of all graphs we consider. The more users join the network, the 
more existing friendship links are uncovered. In Figure~\ref{fig:degree} we
show a loglog-scaled plot of the same data: with the small set of points at our
disposal, it is difficult to draw reliable conclusions, but we are
not always observing the power-law behaviour suggested in~\cite{LKFGEDSD}: see,
for instance, the change of the slope for the \texttt{us} graph.\footnote{We remind the
reader that on a log-log plot almost anything ``looks like'' a straight line. The
quite illuminating examples shown in~\cite{LADTTSFG}, in particular, show that
goodness-of-fit tests are essential.}

\begin{figure}
\centering
\includegraphics[scale=.7]{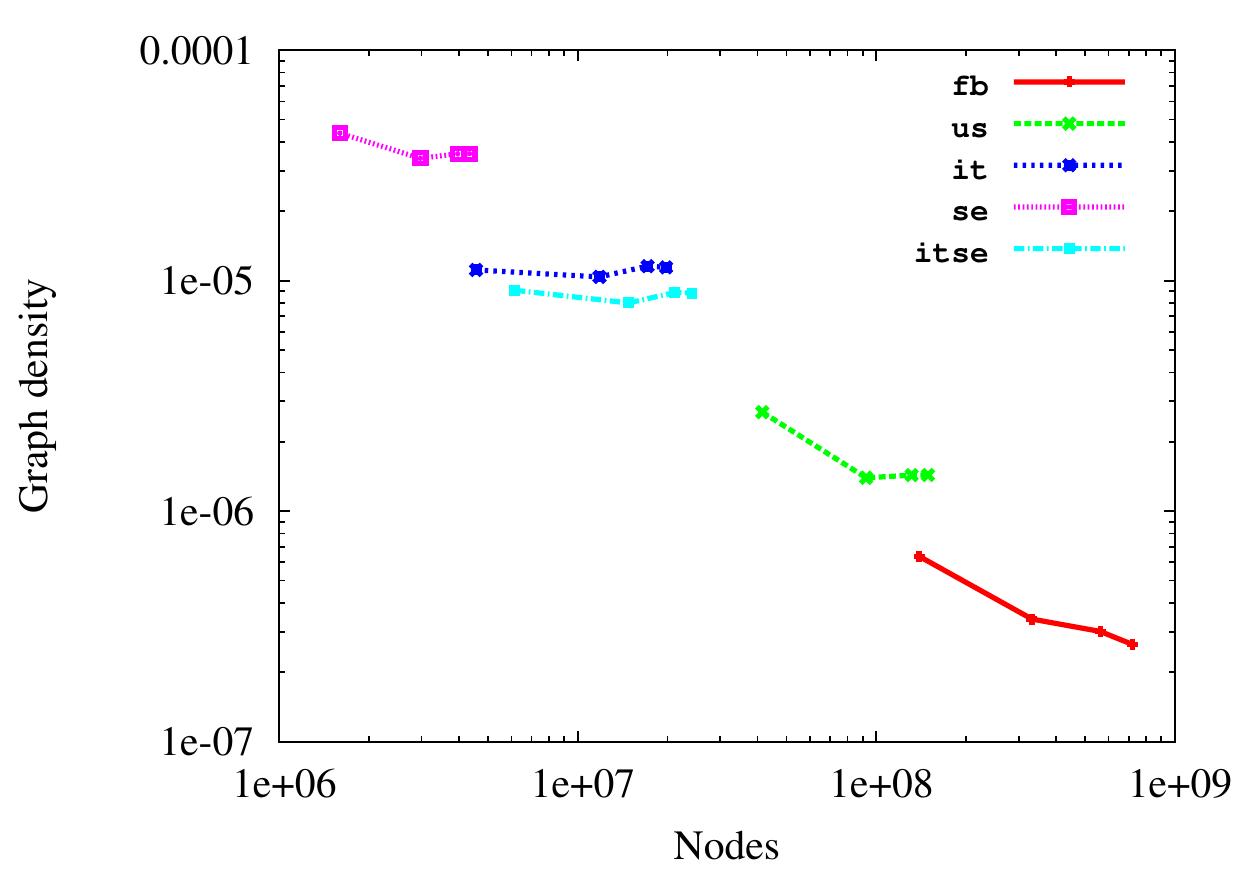}
\caption{\label{fig:density} A plot correlating number of nodes to 
graph density (for the graph from 2009 on).}
\end{figure}

The \emph{density} of the network, on the contrary, decreases.\footnote{We
remark that the authors of~\cite{LKFGEDSD} call \emph{densification} the increase of the average degree, in contrast
with established literature in graph theory, where \emph{density} is the fraction of edges with respect to all
possible edges (e.g., $2m/(n(n-1))$). We use ``density'', ``densification'' and
``sparsification'' in the standard sense.}
In Figure~\ref{fig:density} we plot the density (number of edges
divided by number of nodes) of the graphs against the number of nodes (see also
Table~\ref{tab:density}). There is some initial alternating behaviour, but on
the more complete networks (\texttt{fb} and \texttt{us}) the trend in
sparsification is very evident.
 
Geographical concentration, however, increases density: in
Figure~\ref{fig:density} we can see the lines corresponding to our regional graphs clearly ordered by
geographical concentration, with the \texttt{fb} graph in the lowest
position.

\subsection{Average distance}

The results concerning average distance\footnote{The data we
report is about the average distance \emph{between reachable pairs}, for which
the name \emph{average connected distance} has been proposed~\cite{BKMGSW}. This
is the same measure as that used by Travers and Milgram in~\cite{TMESSWP}. We
refrain from using the word ``connected'' as it somehow implies a bidirectional
(or, if you prefer, undirected) connection. The notion of average distance
between all pairs is useless in a graph in which not all pairs are reachable, as
it is necessarily infinite, so no confusion can arise.} are displayed in
Figure~\ref{fig:avgdist} and Table~\ref{tab:avg}. The average
distance\footnote{In some previous literature (e.g., \cite{LKFGEDSD}), the
$90\%$ percentile (possibly with some interpolation) of the distance
distribution, called \emph{effective diameter}, has been used in place of the
average distance. Having at our disposal tools that can compute easily the
average distance, which is a parameterless, standard feature of the distance
distribution that has been used in social sciences for decades, we prefer to
stick to it. Experimentally, on web and social graphs the average distance is
about  two thirds of the effective diameter plus one~\cite{BRVH}.} on the
Facebook current graph is $4.74$.\footnote{Note that both Karinthy and Guare had
in mind the \emph{maximum}, not the \emph{average} number of degrees, so they
were actually upper bounding the diameter.} Moreover, a closer look at the
distribution shows that $92\%$ of the reachable pairs of individuals are at
distance five or less.

We note that both on the \texttt{it} and \texttt{se}
graphs we find a significantly lower, but similar value. We interpret this
result as telling us that the average distance is actually dependent on the
geographical closeness of users, more than on the actual size of the network.
This is confirmed by the higher average distance of the \texttt{itse} graph.

During the fastest growing years of Facebook our graphs show a quick decrease in
the average distance, which however appears now to be stabilizing. This is not
surprising, as ``shrinking diameter'' phenomena are always observed when a large
network is ``uncovered'', in the sense that we look at larger and larger
induced subgraphs of the underlying global human network. At the same time, as
we already remarked, density was going down steadily. We thus see the
small-world phenomenon fully at work: a smaller fraction of arcs connecting the
users, but nonetheless a lower average distance.

\begin{table*}[ht]
\begin{center}
\begin{tabular}{|c|r|r|r|r|r|}
	\hline
	& \multicolumn{1}{c|}{\texttt{it}} & \multicolumn{1}{c|}{\texttt{se}} & \multicolumn{1}{c|}{\texttt{itse}} & \multicolumn{1}{c|}{\texttt{us}} & \multicolumn{1}{c|}{\texttt{fb}} \\
	\hline
			 2007  &  65\% (11) &  64\% (6) &  67\% (9) &  95\% (5) &  91\% (5) \\
		 2008  &  77\% (7) &  93\% (5) &  77\% (5) &  83\% (5) &  91\% (6) \\
		 2009  &  90\% (5) &  96\% (5) &  75\% (5) &  86\% (5) &  94\% (6) \\
		 2010  &  98\% (5) &  97\% (5) &  91\% (5) &  91\% (5) &  97\% (6) \\
		 2011  &  90\% (4) &  86\% (4) &  95\% (5) &  97\% (5) &  89\% (5) \\
		 current  &  88\% (4) &  86\% (4) &  97\% (5) &  97\% (5) &  91\% (5) \\

	\hline
\end{tabular}
\end{center}
\caption{\label{tab:perc}Percentage of reachable pairs within the ceiling
of the average distance (shown between parentheses).}
\end{table*}

To make more concrete the ``degree of separation'' idea, in
Table~\ref{tab:perc} we show the percentage of reachable pairs \emph{within the ceiling of the
average distance} (note, again, that it is the percentage relatively to the
reachable pairs): for instance, in the current
Facebook graph $92\%$ of the pairs of reachable users are within distance
five---four degrees of separation.

\subsection{Spid}

The \emph{spid} is the \emph{index of dispersion} $\sigma^2/\mu$
(a.k.a.~\emph{variance-to-mean ratio}) of the distance distribution. Some of the
authors proposed the spid~\cite{BRVH} as a measure of the ``webbiness'' of a
social network. In particular, networks with a spid larger than one should be
considered ``web-like'', whereas networks with a spid smaller than one should be
considered ``properly social''. We recall that a distribution is called under-
or over-dispersed depending on whether its index of dispersion is smaller or
larger than 1 (e.g., variance smaller or larger than the average distance), so a
network is considered properly social or not depending on whether its distance distribution
is under- or over-dispersed.

The intuition behind the spid is that ``properly social'' networks strongly
favour short connections, whereas in the web long connection are not uncommon.
As we recalled in the introduction, the starting point of the paper was the
question ``What is the spid of Facebook''? The answer, confirming the data we
gathered on different social networks in~\cite{BRVH}, is shown in
Table~\ref{tab:spid}. With the exception of the highly 
disconnected regional networks in 2007--2008 (see Table~\ref{tab:reachable}),
the spid is well below one.

\begin{figure}
\centering
\includegraphics[scale=.4]{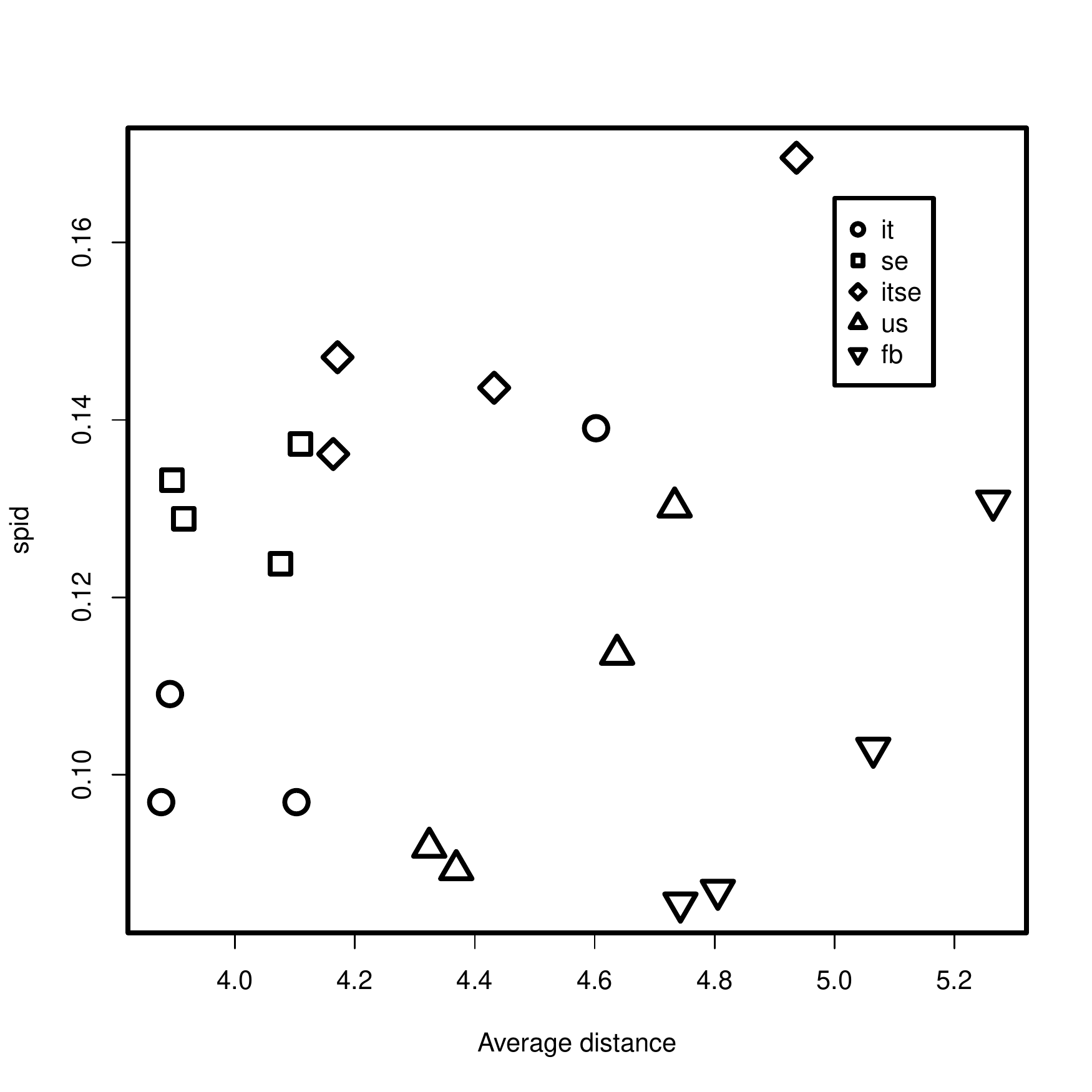}
\caption{\label{fig:avgspid}A scatter plot showing the (lack of) correlation
between the average distance and the spid.}
\end{figure}

Interestingly, across our collection of graphs we can confirm that there is
in general little correlation between the average distance and the spid:
Kendall's $\tau$ is $-0.0105$; graphical evidence of this fact can be seen in 
the scatter plot shown in Figure~\ref{fig:avgspid}. 

\begin{figure}
\centering
\includegraphics[scale=.4]{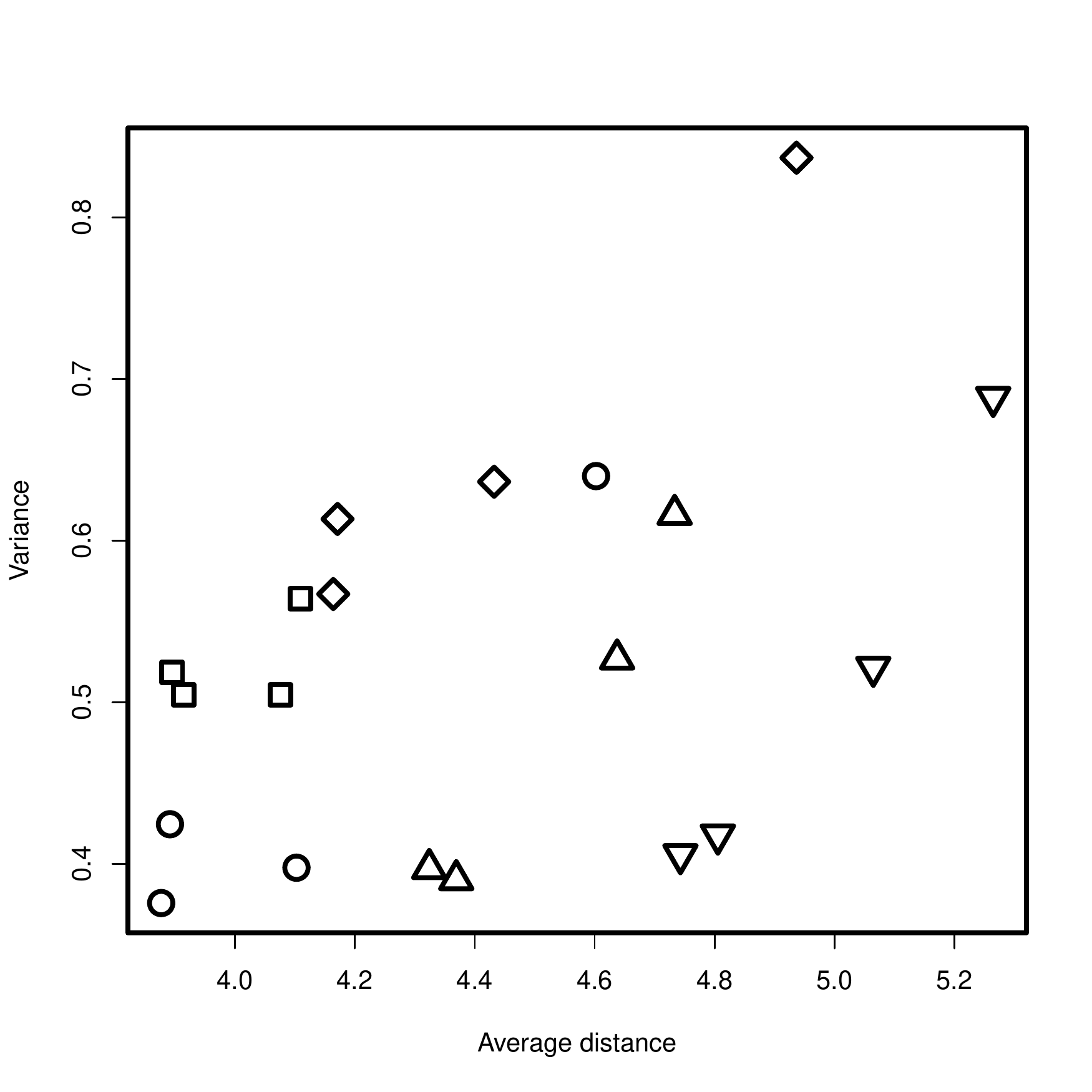}
\caption{\label{fig:avgvar}A scatter plot showing the mild correlation
between the average distance and the variance.}
\end{figure}

If we consider points associated with a single network, though, there appears to
be some correlation between average distance and spid, in particular in the more
connected networks (the values for Kendall's $\tau$ are all above $0.6$, except
for \texttt{se}). However, this is just an artifact, as the correlation between
spid and average distance is \emph{inverse} (larger average distance, smaller
spid). What is happening is that in this case the variance (see
Table~\ref{tab:var}) is changing in the same direction: smaller average
distances (which would imply a larger spid) are associated with smaller
variances. Figure~\ref{fig:avgvar} displays the mild correlation between average
distance and variance in the graphs we analyse: as a network gets tighter, its
distance distribution also gets more concentrated.

% SEBA: does it show variations of the datasets?
% PAOLO: eh? obseve that country-limited graphs tend to be less ``social'' than
% fb, and that fb is getting more and more social in time

\subsection{Diameter}

HyperANF cannot provide exact results about the diameter: however, the number of
steps of a run is necessarily a lower bound for the diameter of the graph
(the set of registers can stabilize before a number of iterations equal to the
diameter because of hash collisions, but never after). While there are no
statistical guarantees on this datum, in Table~\ref{tab:diameter} we report these
maximal observations as lower bounds that differ significantly between regional graphs 
and the overall Facebook graph---there are people that are significantly 
more ``far apart'' in the world than in a single nation.\footnote{Incidentally, 
as we already remarked, this is the measure that Karinthy and Guare 
actually had in mind.}

To corroborate this information, we decided to also approach the problem of
computing the exact diameter directly, although it is in
general a daunting task: for very large graphs matrix-based algorithms are simply not feasible in space, 
and the basic algorithm running $n$ breadth-first visits is not feasible in time.
We thus implemented a highly parallel version of the iFUB (iterative Fringe
Upper Bound) algorithm introduced  in~\cite{CGHCDRWUG} (extending the
ideas of~\cite{CGIFDRWG,MLHFCETBDMG}) for undirected graphs.

The basic idea is as follows: consider some node $x$, and find (by
a breadth-first visit) a node $y$ farthest from $x$. Find now a node $z$ farthest
from $y$: $d(y,z)$ is a (usually very good) lower bound on the diameter, and
actually it \emph{is} the diameter if the graph is a tree (this is the ``double sweep'' algorithm).

We now consider a node $c$ halfway between $y$ and $z$: such a node is ``in the
middle of the graph'' (actually, it would be a \emph{center} if the graph was a
tree), so if $h$ is the eccentricy of $c$ (the distance of the farthest node
from $c$) we expect $2h$ to be a good upper bound for the diameter.

If our upper and lower bound match, we are finished. Otherwise, we consider the
\emph{fringe}: the nodes at distance exactly $h$ from $c$. Clearly, if $M$ is
the maximum of the eccentricities of the nodes in the fringe,
$\max\{\,2(h-1),M\,\}$ is a new (and hopefully improved) upper bound, and $M$
is a new (and hopefully improved) lower bound. We then iterate the process by
examining fringes closer to the root until the bounds match.

Our implementation uses a multicore breadth-first visit: the queue of nodes at
distance $d$ is segmented into small blocks handled by each core. At the end of
a round, we have computed the queue of nodes at distance $d+1$. Our
implementation was able to discover the diameter of the current \texttt{us}
graph (which fits into main memory, thanks to LLP compression) in about twenty
minutes. The diameter of Facebook required ten hours of computation of a machine
with 1TiB of RAM (actually, 256GiB would have been sufficient, always because
of LLP compression).

The values reported in Table~\ref{tab:diameter} confirm what we discovered using
the approximate data provided by the length of HyperANF runs, and suggest that
while the distribution has a low average distance and it is quite concentrated,
there are nonetheless (rare) pairs of nodes that are much farther apart. We
remark that in the case of the current \texttt{fb} graph, the diameter of the
giant component is actually \emph{smaller} than the bound provided by the
HyperANF runs, which means that long paths appear in small (and likely very
irregular) components.

\subsection{Precision}
\label{sec:precision}

As already discussed in~\cite{BRVH}, it is very difficult to obtain strong
theoretical bounds on data derived from the distance distribution. The problem
is that when passing from the neighbourhood function to the distance distribution, the
relative error bound becomes an \emph{absolute} error bound: since
the distance distribution attains very small values (in particular in its tail),
there is a concrete risk of incurring significant errors when computing the
average distance or other statistics. On the other hand, the distribution of derived
data is extremely concentrated~\cite{BRVH}.

\begin{figure}
\centering
\includegraphics[scale=.5]{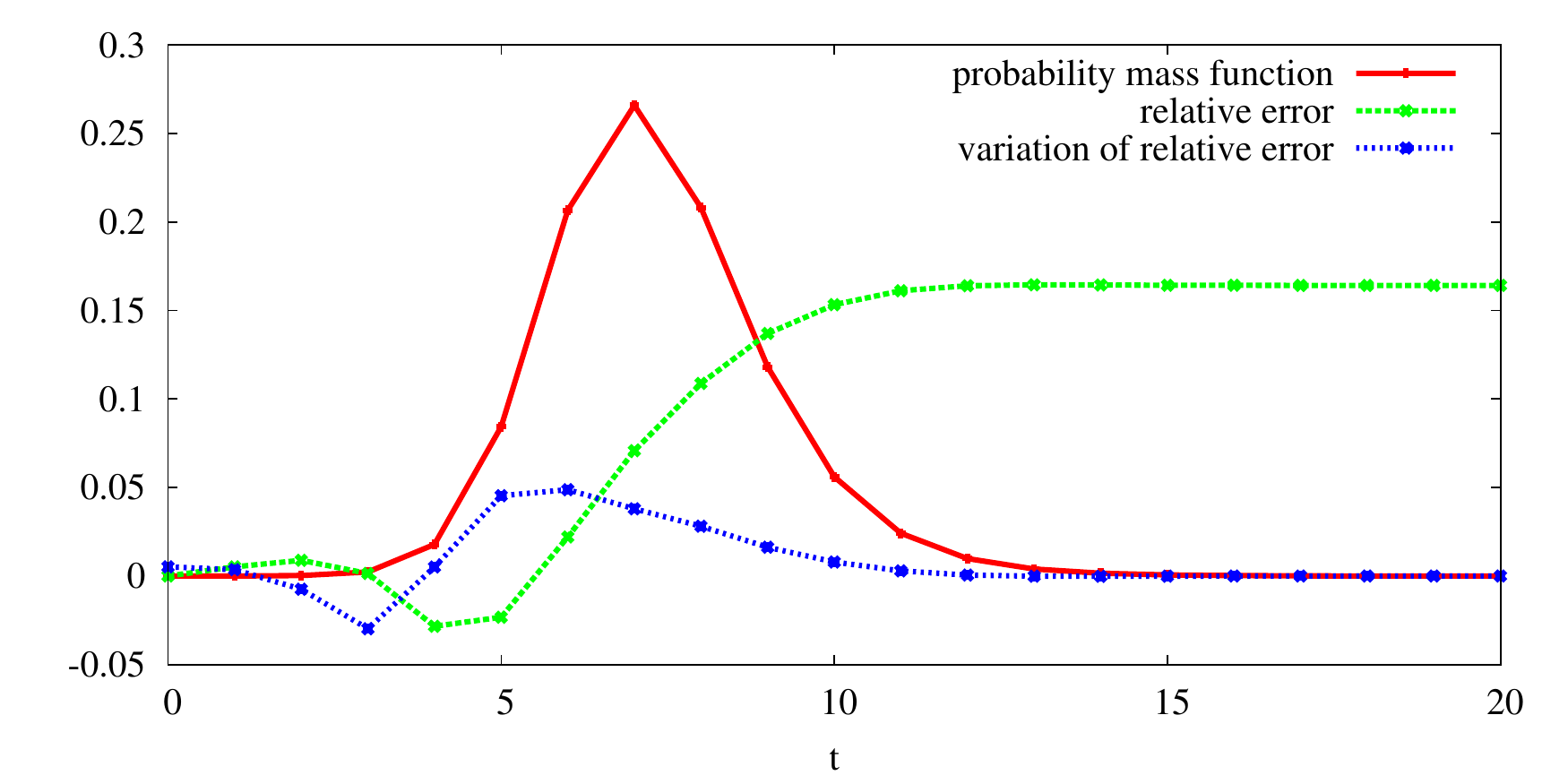}
\caption{\label{fig:evol}The evolution of the relative error in a HyperANF computation
with relative standard deviation $9.25\%$ on a small social network
(\texttt{dblp-2010}).}
\end{figure}

There is, however, a clear empirical explanation of the
unexpected accuracy of our results that is evident from an analysis of the
evolution of the empirical relative error of a run on a social network. 
We show an example in Figure~\ref{fig:evol}.
  \begin{itemize}
    \item In the very first steps, all counters contain essentially disjoint
    sets; thus, they behave as \emph{independent random variables}, and under
    this assumption their relative error should be significantly smaller than
    expected: indeed, this is clearly visible from Figure~\ref{fig:evol}.
    \item In the following few steps, the distribution reaches its highest
    value. The error oscillates, as counters are now significantly dependent
    from one another, but in this part the \emph{actual value of the
    distribution is rather large}, so the absolute theoretical error turns out to be rather
    good.
    \item Finally, in the tail each counter contains a very large
    subset of the reachable nodes: as a result, all counters behave in
    a similar manner (as the hash collisions are essentially the same for every
    counter), and the relative error stabilises to an almost fixed value.
    Because of this stabilisation, \emph{the relative error on the neighbourhood
    function transfers, in practice, to a relative error on the distance
    distribution}. To see why this happen, observe the behaviour of the
    \emph{variation} of the relative error, which is quite erratic initially,
    but then converges quickly to zero. The variation is the only part of the
    relative error that becomes an absolute error when passing to the distance
    distribution, so the computation on the tail is much more accurate than
    what the theoretical bound would imply.
  \end{itemize}
We remark that our considerations remain valid for any diffusion-based
algorithm using approximate, statistically dependent counters (e.g.,
ANF~\cite{PGFANF}).

\section{Conclusions}

In this paper we have studied the largest electronic social network ever created
($\approx721$ million active Facebook users and their $\approx 69$
billion friendship links) from several viewpoints.

First of all, we have confirmed that layered labelled
propagation~\cite{BRSLLP} is a powerful paradigm for increasing locality of a
social network by permuting its nodes. We have been able to compress the
\texttt{us} graph at $11.6$ bits per link---56\% of the
information-theoretical lower bound, similarly to other, much smaller
social networks.

We then analysed using HyperANF the complete Facebook graph and 29 other graphs
obtained by restricting geographically or temporally the links involved.
We have in fact carried out the largest Milgram-like
experiment ever performed. The average distance of Facebook is $4.74$, that is,
$3.74$ ``degrees of separation'', prompting the title of this paper. The spid of
Facebook is $0.09$, well below one, as expected for a social network.
Geographically restricted networks have a smaller average distance, as it happened in Milgram's original experiment. 
Overall, these results help paint the picture of what the Facebook social graph looks like.
As expected, it is a small-world graph, with short paths between many pairs of
nodes.  However, the high degree of compressibility and the study of
geographically limited subgraphs show that geography plays a huge role in
forming the overall structure of network.  Indeed, we see in this study, as
well as other studies of Facebook~\cite{BSMGEO} that, while the world is
connected enough for short paths to exist between most nodes, there is a high
degree of locality induced by various externalities, geography chief amongst
them, all reminiscent of the model proposed in ~\cite{KleNSW}.
%We also observe, over time, a quite erratic variation in density, 
%but the relatively restricted time frame we examine makes drawing 
%stringent conclusions impossible.

When Milgram first published his results, he in fact offered two opposing
interpretations of what ``six degrees of separation'' actually meant. On the one
hand, he observed that such a distance is considerably smaller than what one
would naturally intuit. But at the same time, Milgram noted that this result
could also be interpreted to mean that people are on average six ``worlds
apart'': ``When we speak of five\footnote{Five is the median of the number of
intermediaries reported in the first paper by Milgram~\cite{MilSWP}, from which
our quotation is taken. More experiments were performed with
Travers~\cite{TMESSWP} with a slightly greater average, as reported in
Section~\ref{sec:related}.} intermediaries, we are talking about an enormous
psychological distance between the starting and target points, a distance which
seems small only because we customarily regard `five' as a small manageable
quantity. We should think of the two points as being not five persons apart, but
`five circles of acquaintances' apart---five `structures'
apart.''~\cite{MilSWP}. From this gloomier perspective, it is reassuring to see
that our findings show that people are in fact only four world apart, and not
six: when considering another person in the world, a friend of your friend knows
a friend of their friend, on average.

\iffalse
\hyphenation{ Vi-gna Sa-ba-di-ni Kath-ryn Ker-n-i-ghan Krom-mes Lar-ra-bee
  Pat-rick Port-able Post-Script Pren-tice Rich-ard Richt-er Ro-bert Sha-mos
  Spring-er The-o-dore Uz-ga-lis }

\fi

\bibliography{biblio,fb}

\end{document}